\documentclass[conference,a4paper]{IEEEtran}
\pdfoutput=1

\usepackage{cite}
\usepackage{amsmath,amssymb,amsfonts}
\usepackage{algorithmic}
\usepackage{graphicx}
\usepackage{textcomp}
\usepackage{xcolor}
\usepackage{empheq}
\usepackage{algorithm}
\usepackage{algorithmic}
\usepackage{subfigure}
\usepackage{stfloats}
\usepackage{verbatim}
\def\BibTeX{{\rm B\kern-.05em{\sc i\kern-.025em b}\kern-.08em
    T\kern-.1667em\lower.7ex\hbox{E}\kern-.125emX}}

\begin{document}

\title{Optimal Deployment and Operation of Robotic Aerial 6G Small Cells with Grasping End Effectors \\
}

\author{\IEEEauthorblockN{Yuan Liao and Vasilis Friderikos}
\IEEEauthorblockA{Center of Telecommunication Research, 
King's College London,
London, U.K. \\
E-mail: \{yuan.liao, vasilis.friderikos\} @kcl.ac.uk}
}

\maketitle

\begin{abstract}
Although airborne base stations (ABSs) mounted on drones show a significant potential to enhance network capacity and coverage due to their flexible deployment, the system performance is severely limited by the endurance of the on-board battery. To overcome this key shortcoming, we are exploring robotic airborne base station (RABS) with energy neutral grasping end-effectors able to autonomously perch at tall urban landforms. This paper studies the optimal deployment (fly to another grasping location or remain in the same one) and operation (active or sleep at an epoch) of RABS based on the spatio-temporal characteristics of underlying traffic demand from end-users. Specifically, an integer linear programming (ILP) is formulated by exploiting the coupling between these two decisions, that is, the RABS only needs to visit the locations where it is active. A Lagrangian heuristic algorithm is then proposed by exploiting the totally unimodular structure of the ILP formulation. A wide set of numerical investigations reveal that thanks to its mobility, a single robotic aerial small cell is able to outperform five (5) fixed small cells in terms of served user generated traffic within a 16 to 41 hours period. 
\end{abstract}


\section{Introduction}

\begin{figure}[!t]
	\centering
	\includegraphics[width=0.4\textwidth]{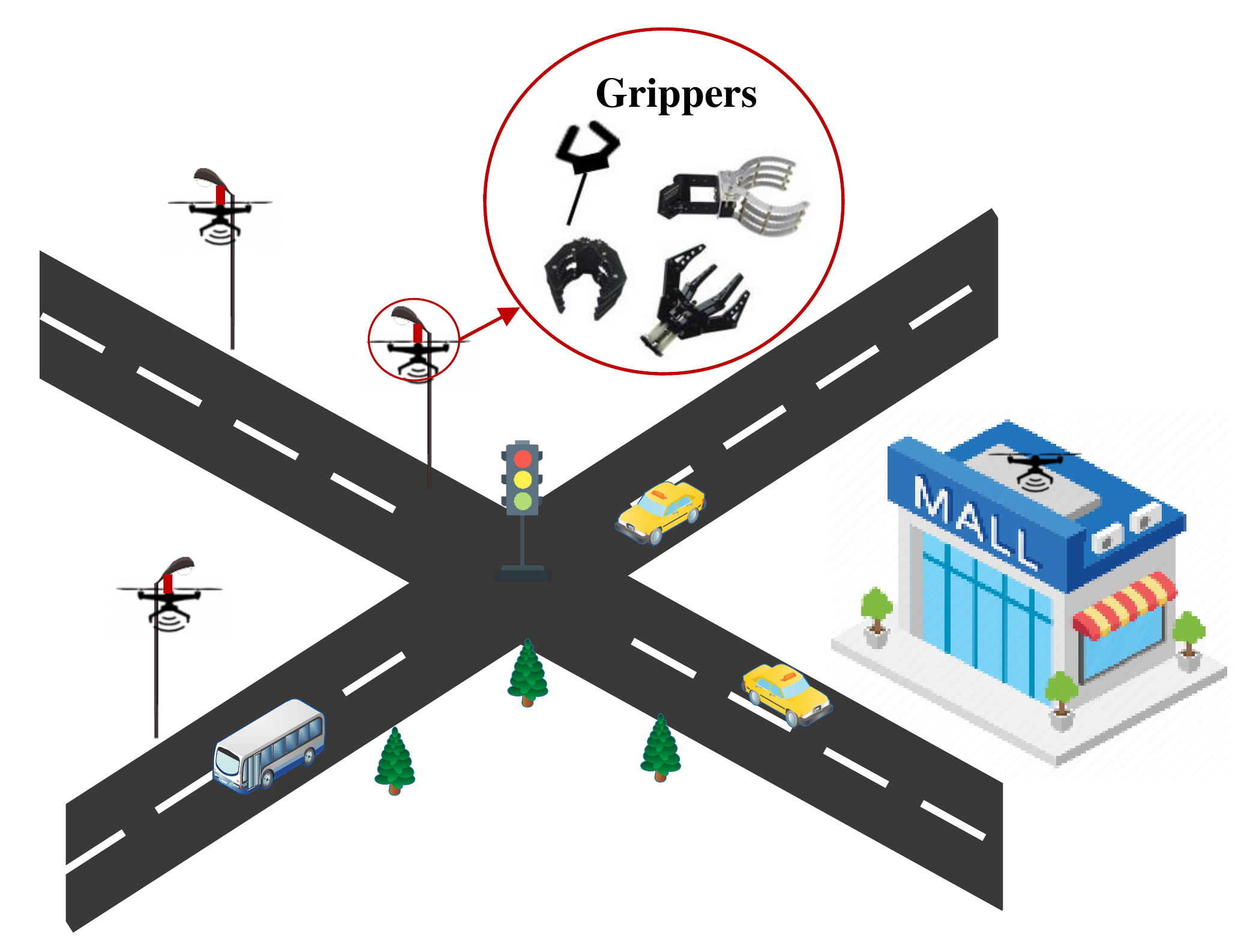}
	\caption{Illustrated use case of airborne 6G microcells with grasping end effectors (RABS). }
	\label{fig:toy}
\end{figure}

Airborne base stations (ABSs) mounted on aerial platforms such as drones are expected to play a significant role in next generation cellular networks, aka 6G, due to their inherent high flexibility and controllable 3D mobility \cite{zeng2019accessing}. However, one of the key limitations when deploying ABSs is the limited battery capacity of drones which immensely curtails the time to act as a small cell. Recently, robotic airborne base stations (RABSs) with grasping capabilities \cite{friderikos2021airborne} or landing based small cells \cite{landedUAV} have been proposed as a mean to provide efficient cell network densification and/or increased network coverage. Advances in grasping capabilities \cite{zhang2020compliant,nedungadi2019design} allow the deployment of robotic small cells that attach autonomously in lampposts (or other tall urban landforms) via energy neutral grasping to act as small cells for multiple hours \cite{friderikos2021airborne}. Therefore, due to their prolonged time availability, RABS might have a significant role to play as 6G  mmWave (or sub-THz) small cells. The use of such high frequency bands will be inevitably required in 6G networks in order to provide 1Gbps support per user to enable support for novel applications with multi modalities such as for example immersive augmented reality and holographic communications. Furthermore, compared to the nominal fixed small cells, RABSs have further degrees of freedoms that allow them not only to perform advanced sleep mode operation but also to change their location, by grasping for example to different lamppost, by exploring (in real-time)  the spatial-temporal dynamics of the traffic in the network \cite{xu2016understanding}. A typical envisioned deployment of RABSs is illustrated in Fig. \ref{fig:toy} where RABSs grasp at roadside lamppost in an urban environment. 
 
A number of efforts are made to overcome the ABSs endurance issue by developing novel serving protocols and ABS prototypes. A novel serving scheme is developed in \cite{xu2018overcoming}, that is, ABS first offloads the files to a subset of users that cache all the files cooperatively, then each user can receive any file from its nearest neighbor that has cached the file via device-to-device communications. In \cite{tuan2021mpc}, an ABS is powered by solar when serving the two-way communication between several pairs of users. In \cite{wu2020fso}, free space optics is applied to transmit both energy and data streams simultaneously between a macro BS and an ABS. Our previous work \cite{friderikos2021airborne} provides a specific introduction of RABS prototype. Another closely related case are tethered drones acting as base stations \cite{kishk2020aerial}. Tethered drones can operate without charging, but their range is limited/fixed to their base unit (which can at the ground or a rooftop). An example of such type of flying base station is ATT’s Flying Cell on Wings (Flying COW)\footnote{When COWs Fly: ATT Sending LTE Signals from Drones, Febr. 2017, www.about.att.com/innovationblog/cows\_fly
} platform which has been designed to provide 4G based network coverage from the sky to ground end users during a disaster event or to increase capacity in case of a major event. Also, F-cell from Nokia is a another closely related technology\footnote{ www.nokia.com/about-us/news/releases/2016/10/03/f-cell-technology-from-nokia-bell-labs-revolutionizes-small-cell-deployment-by-cutting-wires-costs-and-time/}. F-cell is a small cell which is  carried by a drone and is left in rooftops. The key difference is that RABS posses robotic end effectors that can grasp in different urban landforms whereas F-cell is restricted to only landing based operation. 

Sleep mode technique has attracted significant attention in the area of green cellular networking. In \cite{marsan2012multiple}, the optimal decision of sleep mode is identified as a function of the daily traffic pattern. The work in \cite{saker2012optimal} aims to minimize the energy consumption of a heterogeneous network (HetNet) by controlling the sleep mode of BSs adaptively while satisfying the quality of service (QoS). A cross-layer optimization framework and a deep reinforcement learning based method are studied in \cite{cai2013cross} and \cite{ye2019drag} to solve the energy efficiency problem in the HetNet, respectively. The trade-off between network utility and power consumption is achieved by optimizing the BS association and activation in \cite{shen2017flexible}. A specific survey reviewing sleep mode techniques in green cellular networks can be found in \cite{wu2015energy}.

This paper studies the deployment and operation problem of a battery-limited RABS,  aiming to maximize the served traffic load by optimizing two decisions, i.e. active/sleep mode and fly or remain in the same grasping point. This problem is first formulated as an integer linear programming (ILP) by exploiting the coupling between these two decisions, that is, the RABS only need to visit the locations where it is active. Afterwards, by exploiting the totally unimodular structure of the ILP formulation, a Lagrangian heuristic algorithm is proposed to solve it effectively. Numerical results show that the proposed RABS has a significant gain than fixed small BSs thanks to its mobility, especially when the traffic spatial distribution is highly heterogeneous. 

\section{System Model and Problem Formulation}

Hereafter, we assume $M$ candidate locations distributed in a certain geographical area which can be chosen by RABS for grasping; the locations are denoted by $\mathbf{w}_m \in \mathbb{R}^2, \, m = 1,2,...,M$. In practice, the candidate locations can be selected as the roadside lamppost units or other tall urban landforms suitable for grasping \cite{nedungadi2019design}. Furthermore, the time horizon is discretized into $N$ epochs. Each epoch has the same time duration $\delta$, during which the active/sleep mode of the RABS remains unchanged.


\begin{figure}[!t]
\centering
\subfigure{\includegraphics[width=.4\textwidth]{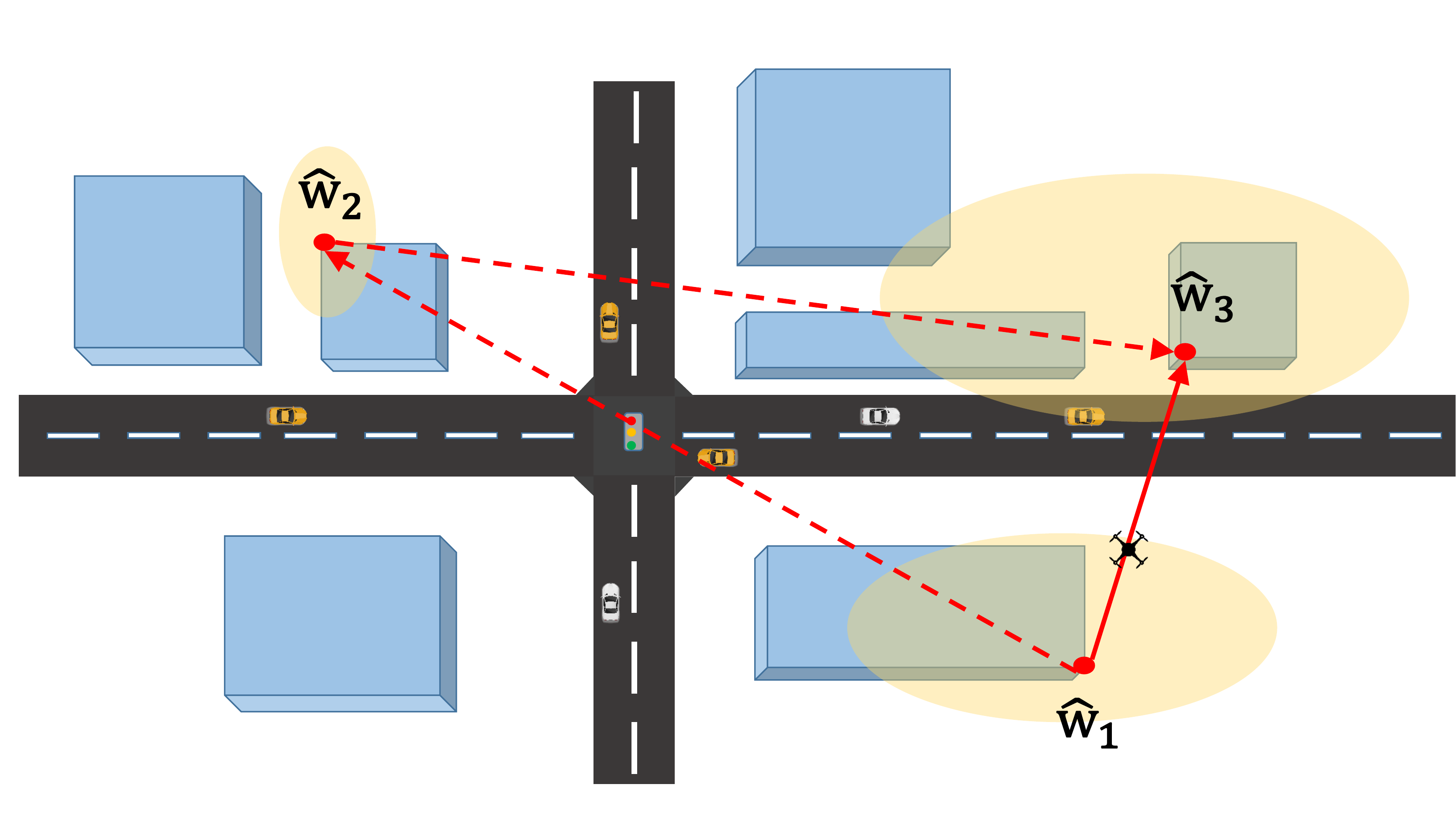}}
\caption{Application scenario: Light blue squares are the city buildings and red dots are the optimal locations for each epoch. Light yellow ellipses represent the traffic demand and its volume is denoted by the area of ellipses. The dotted red line is the RABS flying route without the on-board limitation while the solid red line denotes the flying route of the battery-limited RABS.}
\label{scenario}
\vspace{-0.3cm}
\end{figure}

Similar to \cite{marsan2012multiple,saker2012optimal,cai2013cross,ye2019drag,shen2017flexible}, we assume that the mobile traffic distribution can be predicted accurately. The traffic forecasting and distribution modeling is detailed in section \ref{trafficmodel}. An example scenario when setting $N=3$ is depicted in Fig.\ref{scenario}. According to the spatial traffic distribution at each epoch, we select the optimal location from all candidate locations to place the RABS, shown as red dots in Fig.\ref{scenario} and denoted by $\hat{\mathbf{w}}_n$ for epoch $n$. Without energy limitation, RABS can visit all optimal locations at corresponding epochs and hence achieve the highest traffic load; the flying route to achieve this is shown as the dotted red line in Fig.\ref{scenario}. However, when battery-limited constraints kicks in the optimal strategy, RABS might be to remain at a sleep state at some epochs to save energy and be operational at other epochs where more offered load can be served. In Fig.\ref{scenario}, considering the smaller traffic at epoch 2, denoted by the light yellow ellipses, the relative longer flying distance when visiting $\hat{\mathbf{w}}_2$ is not economical. Thus, the RABS would be active at epochs 1 and 3, and switch to sleep mode at epoch 2. Note that the two operation decisions, i.e., active/sleep mode and change or not the grasping location, are tightly coupled with each other, that is, the RABS only need to visit the location $\hat{\mathbf{w}}_n$ when it is active at epoch $n$. As shown in Fig.\ref{scenario}, since the RABS would be at sleep mode during epochs 2, it would remain grasping at location $\hat{\mathbf{w}}_1$ during the epoch 2 and fly to $\hat{\mathbf{w}}_3$ directly at the beginning of epoch 3. This coupling in decision makings is formulated as a mathematical programming in section \ref{problemformulation}.

\subsection{Spatial-temporal Traffic Modeling }
\label{trafficmodel}

\begin{figure}[!t]
\centering 
\subfigure[Temporal distribution in 24 hour]{\label{temporal_dis} 
\includegraphics[width=.42\textwidth]{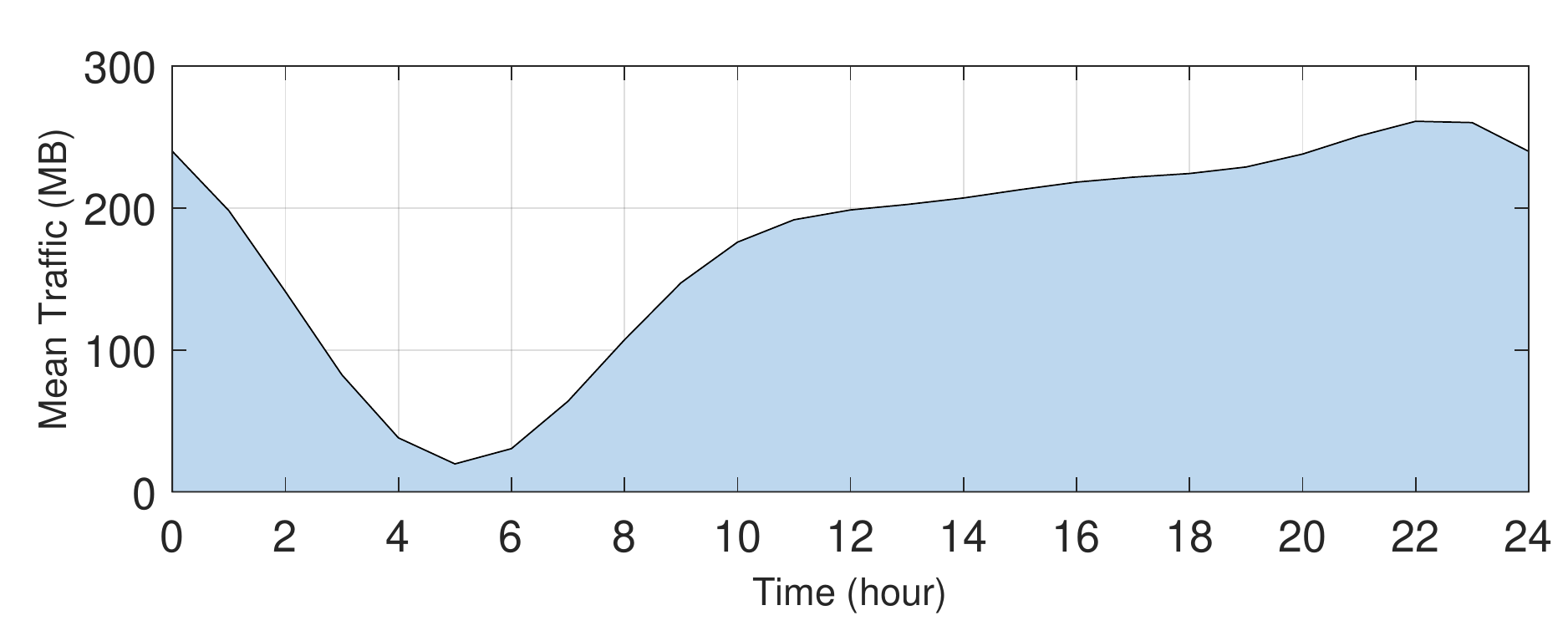}
} 
\subfigure[Spatial distribution at 12 o'clock when $\sigma = 3.6$]{ 
\label{spatial_36}
\includegraphics[width=.225\textwidth]{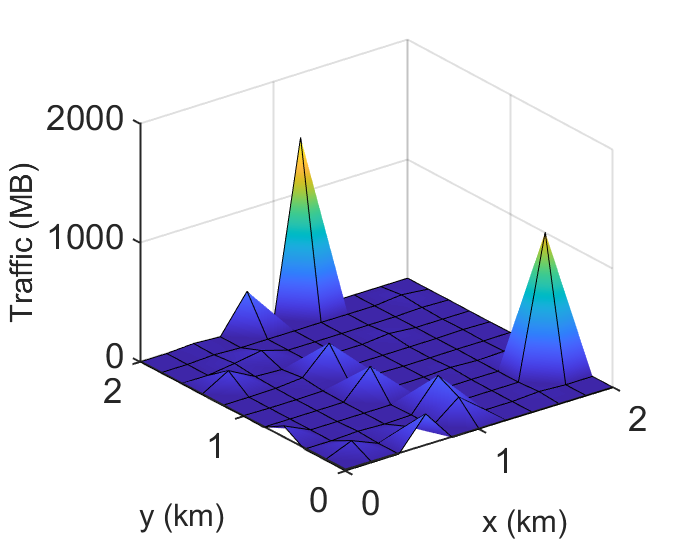}} 
\subfigure[Spatial distribution at 12 o'clock when $\sigma = 1.3$]{ 
\label{spatial_13}
\includegraphics[width=.225\textwidth]{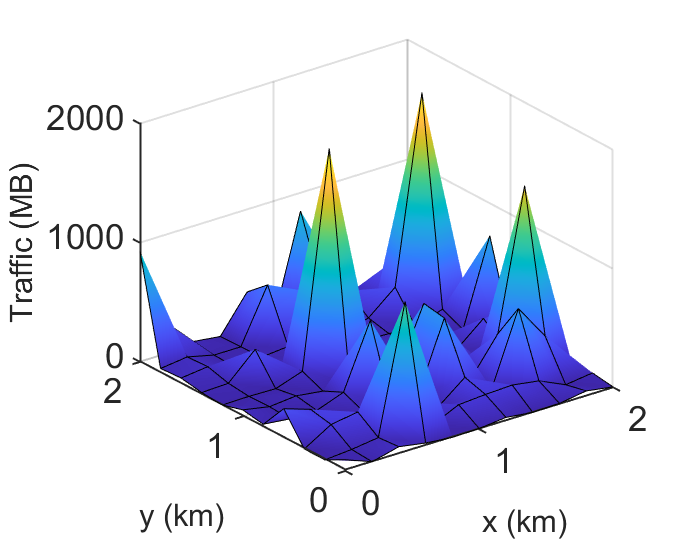}} 
\caption{Visulaization of spatio-temporal traffic distribution}
\label{Trafficdistribution} 
\vspace{-0.3cm}
\end{figure}

In urban areas, the mobile traffic distribution shows a high inhomogeneity in both spatial and temporal domains \cite{xu2016understanding}. However, via machine learning techniques \cite{guo2021can}, the traffic load can be predicted accurately. We utilize the spatial-temporal traffic model proposed in \cite{wang2015approach}. More specifically, the spatial-temporal traffic modeling in \cite{wang2015approach} comprises of two steps. Firstly, we set the epoch duration $\delta = \textit{1 hour}$ and calculate the mean traffic volume in the whole area using the sinusoid superposition model, that is,
\begin{equation}
\label{totaltraffic}
\begin{aligned}
& V(n)  =  173.29 + 89.83 \times \sin{(\frac{\pi}{12}n+3.08}) \\
& + 52.6 \times \sin{(\frac{\pi}{6}n+2.08)} + 16.68 \times \sin{(\frac{\pi}{4}n+1.13)}
\end{aligned} 
\end{equation}
where the three main frequency components, $\frac{\pi}{12}$,  $\frac{\pi}{6}$ and $\frac{\pi}{4}$, show that the traffic volume has a period of 24 hours. Secondly, we generate the predicted traffic value if the RABS is placed at the candidate location $m$ at epoch $n$ using  log-normal distributed samples,
\begin{equation}
\label{lognormal}
\begin{aligned}
V_m(n) = \text{lognrnd}\big(\log(V(n)) - \frac{1}{2}\sigma,\sigma\big),\quad \forall m
\end{aligned} 
\end{equation}
where $\sigma$ is the standard deviation obtained from historical data,  $\text{lognrnd}\big(\log(V(n)) - \frac{1}{2}\sigma,\sigma\big)$ represents the lognormal distribution with mean $\log(V(n)) - \frac{1}{2}\sigma$ and standard deviation $\sigma$. Fig.\ref{Trafficdistribution} illustrates the spatial-temporal traffic model. Comparing Fig.\ref{spatial_36} and Fig.\ref{spatial_13}, the higher the value of $\sigma$ the more inhomogeneous becomes the spatial distribution.

To find the optimal deployment of RABS at each epoch, we select the location with the largest volume from all candidate locations. To this end, we define the largest traffic load at epoch $n$ as $\hat{V}_n \triangleq \max_{m\in \{1,2,...M\}}V_m(n)$, the corresponding index of candidate location as $\hat{m} \triangleq \arg\max_{m\in \{1,2,...M\}}V_m(n)$ and optimal location at epoch $n$ as $\hat{\mathbf{w}}_n \triangleq \mathbf{w}_{\hat{m}}$.

\subsection{RABS Energy Model}
\label{energymodel}

The energy consumed by the RABS is composed of three parts: the propulsion energy for flying, the grasping energy and the communication energy. 

\textit{1) Propulsion Energy:} The propulsion power of a rotary-wing RABS is a function of flying velocity $v$ and given by \cite{zeng2019energy},
\begin{small}
\begin{equation}
\label{flyingpower}
\begin{aligned}
P^{fly} = P_0\Big(1+\frac{3v^2}{U_{tip}^2} \Big) + P_i \Big(\sqrt{1+\frac{v^4}{4v_0^4}} - \frac{v^2}{2v_0^2} \Big)^{1/2} + \frac{1}{2}d_0\rho sAv^3
\end{aligned} 
\end{equation}
\end{small}where $P_0$ and $P_i$ represent blade profile power and induced power, respectively. $U_{tip}$ is the tip speed of the rotor blade. $v_0$ denotes the mean rotor induced velocity when hovering. $d_0$ and $s$ are the fuselage drag ratio and rotor solidity, respectively. $\rho$ and $A$ denote the air density and rotor disc area, respectively. Subsequently, the propulsion energy to fly between two optimal locations, $\hat{\mathbf{w}}_n$ and $\hat{\mathbf{w}}_{n'}$, is given by,
\begin{equation}
\label{flyingenergy}
\begin{aligned}
E^{fly}_{nn'} = \frac{P^{fly} \lVert \hat{\mathbf{w}}_{n'} - \hat{\mathbf{w}}_n  \rVert^2}{v}, \; n' > n
\end{aligned} 
\end{equation}
where $\lVert\cdot\rVert$ denotes the 2-norm for a vector. Notably, $n' > n$ means that the RABS can only fly from the optimal location at epoch $n$ to the optimal location at epoch $n'$ but cannot move in the opposite direction since the time passed.

\textit{2) Grasping Energy:} The grasping power of RABS depends on its size and weight, as well as the type of the gripper \cite{friderikos2021airborne}. Electromagnetic solenoid based grippers can be deemed as suitable for attaching to ferromagnetic surfaces such as lampposts \cite{nedungadi2019design}. 
The grasping energy consumed during an epoch $\delta$ is given by $E^{grasp} = P^{grasp} \delta$, where $P^{grasp}$ is the grasping power.

\textit{3) Communication Energy:} Assuming a micro cell \cite{auer2011much},
the communication energy  for active and sleep mode is given by $E^{active} = (\eta P^{tra} + P^{active})\delta$ and $E^{sleep} = P^{sleep}\delta$, respectively, where $P^{tra}$, $P^{active}$ and $P^{sleep}$ are the transmission power, active and sleep mode power, respectively. $\eta > 0$ is a coefficient reflecting the power amplifier efficiency, feeder loss and other loss factors.

\subsection{Problem Formulation and Analysis}
\label{problemformulation}

According to the illustrated scenario and aforementioned
system model, an ILP problem is formulated in this subsection to maximize the traffic load by operating the RABS mode under the on-board battery constraint. Unlike the ground base station, of which the operator can only decide the active/sleep mode, the operation problem of RABS has two freedom, active/sleep and fly/not. Thus, there are two sets of binary variables are used to formulate these decisions. $x_{n} \in \{0,1\}$ denotes the RABS would be active ($x_{n} = 1 $) or sleep ($x_{n} = 0 $) at epoch $n$, and $y_{nn'} \in \{0,1\}$ indicates whether ($y_{nn'} = 1$) or not ($y_{nn'} = 0$) the RABS flies from $\hat{\mathbf{w}}_n$ to $\hat{\mathbf{w}}_{n'}$. Similar as \eqref{flyingenergy}, we limit $n'>n$ for $y_{nn'}$ to satisfy the order of precedence. 

To exploit the coupling between two decisions illustrated in Fig.\ref{scenario}, we first introduce two pseudo-epochs indexed by $0$ and $N+1$. Then, this coupling relationship can be formulated as,
\begin{subequations}
\begin{empheq}[left={\empheqlbrace\,}]{align}
&  \sum_{n'=n+1}^{N+1}y_{nn'} = x_{n}, \quad n = 0,1,...,N \label{couplingcon1} \\
&  \sum_{n'=0}^{n-1}y_{n'n} = x_{n}, \quad n = 1,...,N,N+1 \label{couplingcon2} 
\end{empheq}
\end{subequations}
where \eqref{couplingcon1} denotes that the RABS would depart from $\hat{\mathbf{w}}_{n}$ ( $\sum_{n'=n+1}^{N+1}y_{nn'} = 1$ ) only when it is active at epoch n ( $x_{n} = 1$ ) while \eqref{couplingcon2} denotes the arriving process. For simplicity of formulation, the settings of two pseudo-epochs are given as,
\begin{subequations}
\begin{empheq}[left={\empheqlbrace\,}]{align}
& E^{fly}_{nN+1} = 0, \quad n = 0,1,...,N \label{pseudo1} \\
& E^{fly}_{0n} = E^{fly}_{1n}, \quad n = 1,2,...,N+1  \label{pseudo2}
\end{empheq}
\end{subequations}
\eqref{pseudo1} denotes that the propulsion energy consumed by flying from any $\hat{\mathbf{w}}_n$ to $\hat{\mathbf{w}}_{N+1}$ is $0$, while \eqref{pseudo2} represents that the propulsion energy of any route departing from $\hat{\mathbf{w}}_0$ is equal to which leaving from $\hat{\mathbf{w}}_1$. Subsequently, the operation problem of RABS can be formulated as,
\begin{subequations}
\begin{align}
\mathrm{(P1):} 
\; &  \max_{\mathbf{X}, \mathbf{Y}} \sum_{n=1}^{N} \hat{V}_n x_{n} \label{Pro1obj} \\
s.t.
\; & \eqref{couplingcon1}-\eqref{couplingcon2} \label{Pro1C1} \\
\; & E( \mathbf{X} , \mathbf{Y}) \leq E^{max}  \label{Pro1C2}\\
\; & x_{n} \in \{0,1\}, \quad \forall n \in  \{0,1,...,N,N+1\} \label{Pro1C3} \\
\; & y_{nn'} \in \{0,1\}, \quad \forall nn' \in  \{nn' \big| \notag \\ 
\; &  \qquad \;n \!\in \{0,..,N\}, n'\!\in \{1,..,N\!+\!1\}, n' \!>\!n\} \footnotemark \label{Pro1C4} 
\end{align}
\end{subequations}where $\mathbf{X} \triangleq \{x_{n}\} $ and $\mathbf{Y} \triangleq \{y_{nn'}\}$ are the sets of variables. $E( \mathbf{X} , \mathbf{Y})$ calculates the total consumed energy and is a function of $\mathbf{X}$ and $\mathbf{Y}$, that is,
\begin{equation}
\label{totalenergy}
\begin{aligned}
E( \mathbf{X} , \mathbf{Y}) \triangleq & \sum_{n=0}^{N}\sum_{n'=n+1}^{N+1}y_{nn'} E^{fly}_{nn'} \!+ \! \sum_{n=1}^{N} x_{n} (E^{active} + E^{grasp}) \\ 
\; & + \sum_{n=1}^{N} (1-x_{n}) (E^{sleep} + E^{grasp})
\end{aligned} 
\end{equation}

The objective function \eqref{Pro1obj} maximizes the served traffic. Constraints in \eqref{Pro1C1} exhibit the coupling relationship between two decisions, that is, the RABS would visit the location $\hat{\mathbf{w}}_n$ only when it is active at epoch $n$. \eqref{Pro1C2} is the energy constraint where $E^{max}$ is the on-board battery capacity. Clearly, (P1) is an ILP and the constraint \eqref{Pro1C2} is a binary knapsack constraint. Thus, (P1) is not easier than the binary knapsack problem, which is a well-known NP-hard problem. According to \eqref{Pro1C3} and \eqref{Pro1C4}, the number of variables in (P1) can be given by $ \big|\mathbf{X}\big| + \big|\mathbf{Y}\big| = (N+2) + (N+2)!/2N!$, where $\lvert\cdot\rvert$ denotes the cardinality of a set. It can be seen that the scale of (P1) grows sharply as $N$ increases. To overcome the curse of dimensionality, we aim to propose an efficient algorithm to capture a high-quality solution of (P1). Before designing the algorithm, we would first do some analysis.

\footnotetext{ Hereafter, unless otherwise specified, $\forall nn'$ is used to represent $\forall nn'\in \! \{nn' \big| n\! \in\! \{0,..,N\}, n' \!\in \!\{1,..,N\!+\!1\}, n'\!>\!n\}$ for simplicity of illustration.} 

\textit{\textbf{Lemma 1:} The parameter matrix of constraints \eqref{couplingcon1}-\eqref{couplingcon1} is totally unimodular.}

\textit{Proof:} See the proposition 2.6 in the section III.1.2 of \cite{wolsey1999integer}.

Relaxing the energy constraint \eqref{Pro1C2}, $z_{LR}(\lambda)$ given by the following problem (LR-P1) is the Lagrangian relaxation of (P1), where $\lambda$ is the nonnegative Lagrange multiplier. 
\begin{subequations}
\begin{align}
\text{(LR-P1):} 
\; &  z_{LR}(\lambda) = \max_{\mathbf{X}, \mathbf{Y}} \sum_{n=1}^{N} \hat{V}_n x_{n} - \lambda \Big( E( \mathbf{X} , \mathbf{Y}) - E^{max} \Big) \label{LRP1obj} \\
s.t.
\; &  \eqref{couplingcon1}-\eqref{couplingcon2}, \eqref{Pro1C3}-\eqref{Pro1C4} \label{LRP1C1}
\end{align}
\end{subequations}
Certainly, $z_{LR}(\lambda)$ provides an upper bound for (P1), the least upper bound can be achieved by solving the following Lagrangian dual problem,
\begin{equation}
\label{Lagrangiandual}
\begin{aligned}
\quad & z_{LD} = \min_{\lambda \geq 0} z_{LR}(\lambda)
\end{aligned} 
\end{equation}

Another normally used relaxation technique of ILP is linear relaxation, in which the integer constraints are relaxed to continuous constraints. The linear relaxation of (P1) is given by,
\begin{subequations}
\begin{align}
\text{(LP-P1):} 
\quad & z_{LP} = \max_{\mathbf{X}, \mathbf{Y}} \sum_{n=1}^{N} \hat{V}_n x_{n}  \label{LRPro1obj} \\
s.t.
\quad & \eqref{couplingcon1} - \eqref{couplingcon2}, \eqref{Pro1C2} \label{LRPro1C1}\\
\quad & 0 \leq x_{n} \leq 1, \quad \forall n  \label{LRPro1C2} \\
\quad & 0 \leq y_{nn'} \leq 1, \quad \forall nn'  \label{LRPro1C3}
\end{align}
\end{subequations}
The following proposition shows that the linear relaxation and Lagrangian dual of (P1) have the equal objective value. This fact is utilized when we select a suitable step size for the subgradient method in section \ref{LagrangianHeuristic}.

\textit{\textbf{Proposition 1:} $z_{LD} = z_{LP}$}

\textit{Proof:} Refer to the corollary 6.6 in section II.3.6 of \cite{wolsey1999integer}.

\section{Lagrangian Heuristic Algorithm}
\label{LagrangianHeuristic}

In this section, based on the preceding analysis, a Lagrangian heuristic algorithm is proposed to overcome the curse of dimensionality in (P1). Specifically, although proposition 1 shows the equality between the objective values of linear relaxation and Lagrangian dual, solving (LP-P1) may not obtain an integer result so that difficult to refine a feasible solution for (P1). We propose a Lagrangian heuristic algorithm in this section, the basic idea of which is to solve the Lagrangian dual \eqref{Lagrangiandual} through the subgradient method, and construct a feasible solution for (P1) through a refinement procedure. Compared with solving the linear programming (LP-P1) directly, we would show that the result captured by Lagrangian relaxation is integer, thus easing the construction of feasible solutions.

\textit{1) Subgradient Method for solving Lagrangian dual \eqref{Lagrangiandual}:} In this step, the subgradient method is utilized to solve the Largrangian dual \eqref{Lagrangiandual}, which is convex and non-smooth. To obtain a subgradient direction for any given $\lambda$, we would first solve the Lagrangian relaxation (LR-P1), which is still an ILP and difficult to solve. Fortunately, Lemma 1 notes that the parameter matrix of \eqref{couplingcon1}-\eqref{couplingcon1} is totally unimodular, and proposition 2.2 in section III.1.2 of \cite{wolsey1999integer} illustrates that the solution of (LR-P1) can be obtained by solving its linear relaxation (P2),
\begin{subequations}
\begin{align}
\text{(P2):} 
\quad &  \max_{\mathbf{X}, \mathbf{Y}} \sum_{n=1}^{N} \hat{V}_n x_{n} - \lambda \Big( E( \mathbf{X} , \mathbf{Y}) - E^{max} \Big) \label{P2obj} \\
s.t.
\quad &  \eqref{couplingcon1}-\eqref{couplingcon1}, \label{Pro2C1} \\
\quad & 0 \leq x_{n} \leq 1, \quad \forall n  \label{Pro2C2} \\
\quad & 0 \leq y_{nn'} \leq 1, \quad \forall nn'  \label{Pro2C3}
\end{align}
\end{subequations}

Accordingly, a normally used subgradient direction for $\lambda$ is given by \cite{fisher1981lagrangian}, 
\begin{equation}
\label{subgradient}
\begin{aligned}
g = E( \mathbf{X} , \mathbf{Y}) - E^{max}
\end{aligned} 
\end{equation}
To minimize $z_{LR}(\lambda)$ shown in \eqref{Lagrangiandual}, $\lambda$ is updated as,
\begin{equation}
\label{updatelambda}
\begin{aligned}
\lambda_{k+1} = [\,\lambda_{k} - \alpha_k g_k\,]^+
\end{aligned} 
\end{equation}
where $[\cdot]^+$ denotes projection onto the positive orthant, $\lambda_k$, $\alpha_k$ and $g_k$ are the Lagrangian multiplier, step size and subgradient at iteration $k$, respectively. Notably, $\alpha_k$ should be chosen carefully to guarantee the convergence,
we utilize the method proposed in \cite{bragin2015convergence},
\begin{subequations}
\begin{empheq}[left={\empheqlbrace\,}]{align}
& \alpha_0 = \frac{z_{LP} - z_{LR}(\lambda_0)}{\lVert g_0 \rVert^2} \label{stepsize1} \\
& \alpha_k = \big(1 - \frac{1}{\beta k^{1-k^{-r}}}\big) \frac{\alpha_{k-1}\lVert g_{k-1} \rVert}{\lVert g_k \rVert} \label{stepsize2} 
\end{empheq}
\end{subequations}
where $\beta \geq 1$ and $0<r<1$ are predefined parameters. The procedure of  subgradient method is summarized as the step 1-6 in Algorithm \ref{LagHeu}.

\textit{2) Construct a feasible solution:} Although the solution of linear programming (P2) is integer, it generally does not satisfy the energy constraint \eqref{Pro1C2}. We propose a greedy method to construct a feasible solution. Based on the result captured by dual problem \eqref{Lagrangiandual}, choose the active epoch with the least traffic load, make this epoch sleep and delete the related flying paths from the RABS route. Repeat this process until the energy constraint \eqref{Pro1C2} is satisfied. The greedy method is shown as the step 7-10 in Algorithm \ref{LagHeu}.

\begin{algorithm}[!t]
\caption{Lagrangian Heuristic Algorithm}
\label{LagHeu}
\begin{algorithmic}[1]
\STATE Obtain $z_{LP}$ by solving the linear programming (LP-P1). \label{substart}
\STATE Initialize $\lambda_0$. Obtain $z_{LR}(\lambda_0)$ and $g_0$ by solving (P2). Initialize the step size through \eqref{stepsize1}. Set $k=1$. 
\REPEAT
\STATE Update the $\lambda_k$ through \eqref{updatelambda}. Obtain $z_{LR}(\lambda_k)$ and $g_k$ by solving linear programming (P2). Update the step size through \eqref{stepsize2}.
\STATE $k = k+1$.
\UNTIL{The iteration index $k$ achieves a threshold $k^{max}$. } \label{stoppingcriteria}
\REPEAT \label{refinestart}
\STATE Choose the active epoch with the least traffic load. Make the RABS sleep at this epoch and delete the related route. 
\UNTIL{Energy constraint \eqref{Pro1C2} is satisfied.}\label{refineend}
\end{algorithmic}
\end{algorithm}

\textit{\textbf{Remark 1:}} (Stopping criteria) To solve the Lagrangian dual \eqref{Lagrangiandual} accurately, the stopping criteria of subgradient method is always set as the gap between $z_{LR}(\lambda_k)$ and $z_{LD}$ is smaller than a threshold. However, in the proposed Algorithm \ref{LagHeu}, whenever the subgradient method in step \ref{substart}-\ref{stoppingcriteria} stops, the refinement procedure in step \ref{refinestart}-\ref{refineend} can always return a feasible solution for problem (P1). Therefore, we set the stopping criteria as the maximum number of iterations to control the running time conveniently. 

\textit{\textbf{Remark 2:}} (Computation complexity) Section 6.6.1 of \cite{ben2001lectures} illustrates that the worst case of solving a linear programming is approximately $\mathcal{O}\big((n^v+n^c)^{1.5}{n^v}^2\big)$, where $n^v$ and $n^c$ are the number of variables and constraints, respectively. Moreover, the refinement procedure \ref{refinestart}-\ref{refineend} would check at most $N$ epochs in the worst case. Therefore, the complexity of Algorithm \ref{LagHeu} is approximately $\mathcal{O}\big(k^{max}\cdot(n^v+n^c)^{1.5}{n^v}^2 + N\big)$, where $n^v = (N+2) + (N+2)!/2N!$ and $n^c = (3N+4) + (N+2)!/2N!$ for linear programming (P2).

\section{Numerical Investigations}
\label{NumericalResults}

\begin{figure*}[ht]
\centering
\begin{minipage}[b]{0.32\textwidth} 
\centering
\includegraphics[width=0.95\textwidth]{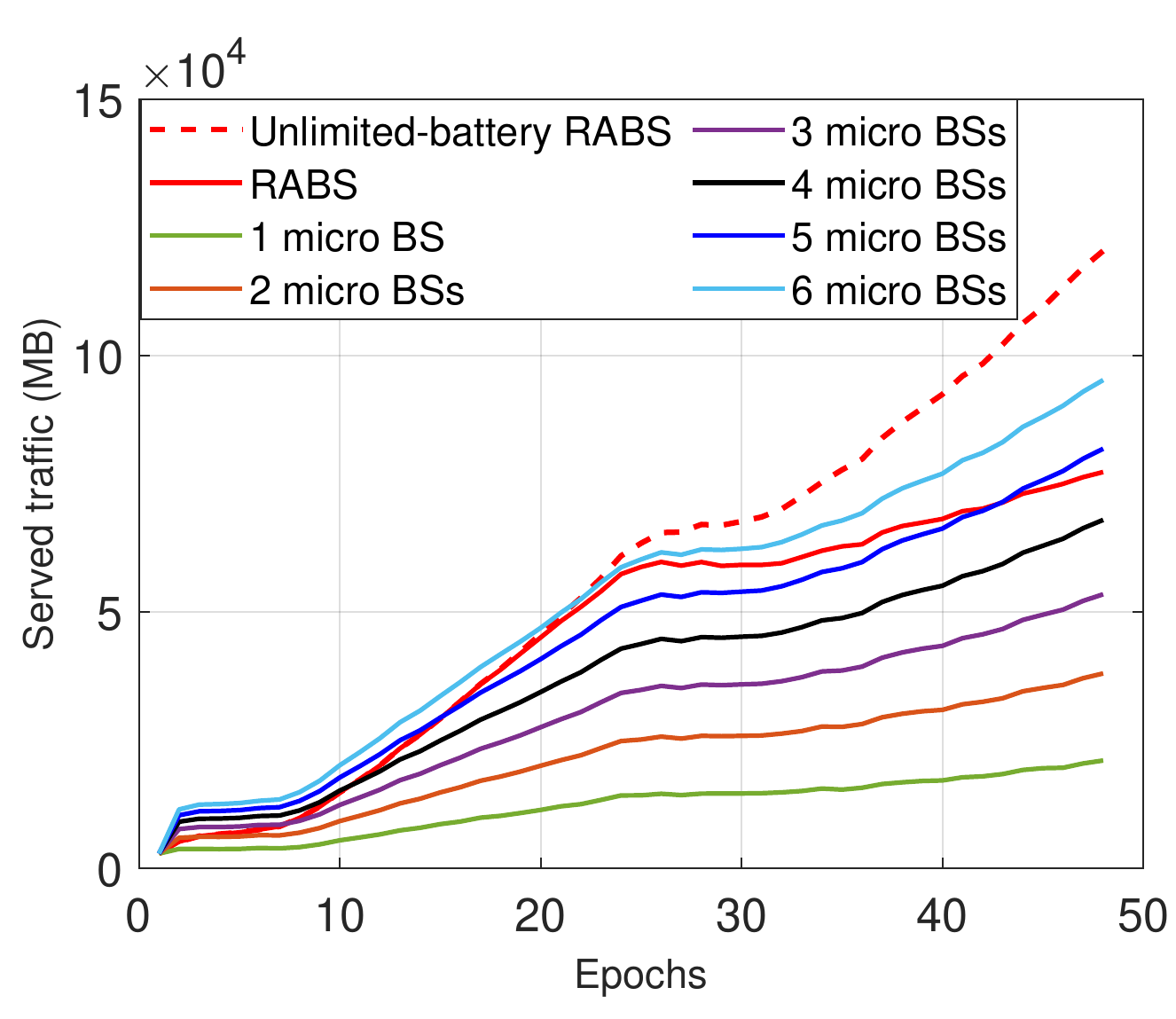} 
\caption{Comparing the RABS with fixed micro BSs}
\label{comparefixedBS}
\end{minipage}
\begin{minipage}[b]{0.32\textwidth}
\centering
\includegraphics[width=0.95\textwidth]{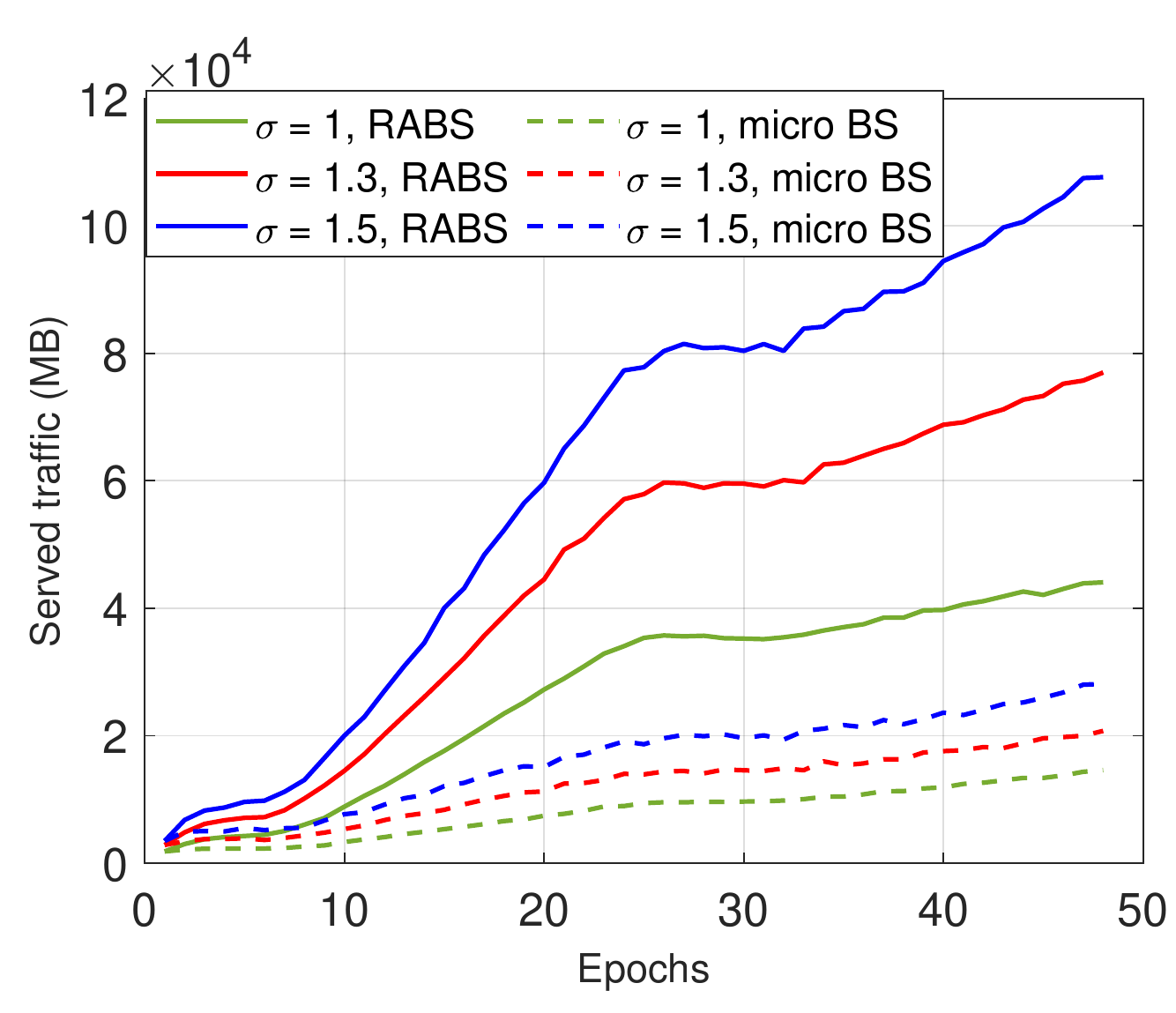}
\caption{Performances of the RABS and the fixed micro BS performances under different $\sigma$}
\label{differentsigma}
\end{minipage}
\begin{minipage}[b]{0.32\textwidth}
\centering
\includegraphics[width=0.9\textwidth]{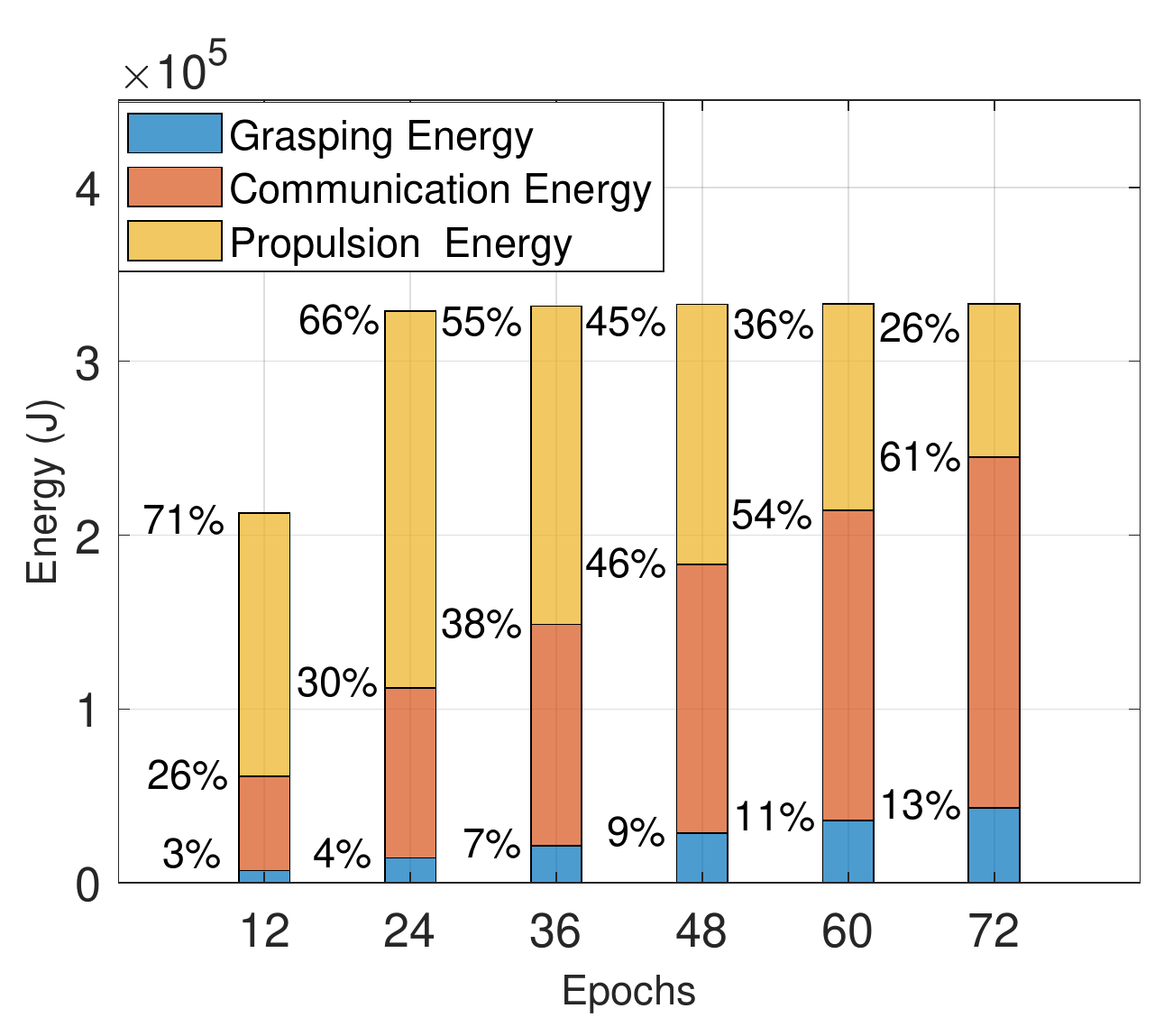}
\caption{Comparison on energy consumption for propulsion, communication and grasping}
\label{energy_components}
\end{minipage}
\end{figure*}

In this section, numerical investigations are presented to evaluate the proposed deployment and operation strategies for RABS. The parameterization settings used hereafter are summarized in Table \ref{TAB para}. Similar as Fig.\ref{spatial_36} and Fig.\ref{spatial_13}, we consider a $2\times2 \, \text{km}^2$ area where 121 candidate locations distributed evenly, that is, $M = 121$. Besides, note that the battery capacity is generally measured in milliampere/hour ($\mathrm{mAh}$) under a certain output voltage. For computing convenience, we calculate the capacity of \textit{Zappers SG4} battery by $15.2 \, \mathrm{V} \times 6100 \, \mathrm{mAh} \times 3.6 = 333792 \, \mathrm{J}$  in this section \cite{friderikos2021airborne}.

\begin{table}[!t]
\centering
\caption{Parameter Settings}
\label{TAB para}
\begin{tabular}{ll|ll}
\hline
Parameter & Value & Parameter & Value\\
\hline
$E^{max}$ & 333792 J  & $v$ & 30 m/s \\
$P^{fly}$ & 356 W \cite{zeng2019energy} & $\eta$ & 2.6 \cite{auer2011much} \\
$P^{grasp}$ & 10 W \cite{nedungadi2019design} & $P^{tra}$ & 6.3 W \cite{auer2011much} \\
$P^{active}$ & 56 W \cite{auer2011much} & $P^{sleep}$ & 39 W \cite{auer2011much} \\
\hline
\end{tabular}
\end{table}

Fig.\ref{comparefixedBS} compares the served traffic by the RABS and different number of micro BSs when setting $N$ from 1 hour to 48 hours. Assuming that the traffic distribution can be predicted accurately, the micro BSs are deployed greedily, e.g. we select the candidate location with the largest traffic volume to deploy the BS when there is one BS available, choose the first two best locations when there are two micro BSs available and so on. We assume that micro BSs are powered by cable thus can provide service all the time. Relaxing the energy constraint \eqref{Pro2C1}, Fig.\ref{comparefixedBS} shows that an 'ideal' RABS can achieve a better traffic offload than six micro BSs when $N\geq24$. However, when the on-board battery capacity is considered, the RABS still has a better performance than five micro BSs when $16 \leq N\leq 41$. Hence, Fig.\ref{comparefixedBS} illustrates the fact that compared with fixed micro BSs, the flexibility of RABS has a significant gain even though its on-board battery is limited.

Fig.\ref{differentsigma} compares the RABS and the micro BS under different value of $\sigma$. Reviewing the traffic distribution \eqref{lognormal} and Fig.\ref{Trafficdistribution}, a larger $\sigma$ shows higher heterogeneity of traffic spatial distribution. Comparing the RABS performances under different $\sigma$, it can be seen that the RABS has a better performance in high heterogeneity scenario. Besides, comparing the micro BS and the RABS performances, Fig.\ref{differentsigma} also shows that the RABS has a larger gain than micro cell in high heterogeneity scenario. For instance, setting $N = 48$, the traffic served by the RABS is 3.8 times more than that by the micro BS when $\sigma = 1.5$, while this gain rate changes to 3.0 when $\sigma = 1.0$.

Fig.\ref{energy_components} investigates three components of energy consumption reviewed in section \ref{energymodel}. It can be seen that the flying and communication procedures consume most energy than grasping. This also explains why RABS has a longer endurance than normal ABS that provides service when flying or hovering in the air thus certainly consumes more energy. Comparing the percentages of propulsion and communication energy, it can be seen that with the increase of serving endurance, the proportion of communication energy becomes larger while the proportion of flight energy gradually decreases. For example, when $N = 24$, the propulsion and communication energy accounts for 66\% and 30\% of the total energy consumption, respectively, while the proportions change to 26\% and 61\% when $N = 72$.

\section{Conclusions}
\label{Conclusions}
RABS with dexterous end effectors able to grasp in an energy neutral manner at different tall urban lanndforms can provide significant degree of flexibility in deploying 6G small cells. RABS allows not only to enter to sleep mode but can also change its perching point based on the spatial-temporal characteristics of the traffic. To this end, an ILP problem has been formulated to optimize the deployment and operation of RABS based on traffic demand, which is solved by the Lagrangian heuristic algorithm effectively. Numerical investigations reveal that the RABS significantly outperform the nominal fixed small cells thanks to its mobility, especially when the traffic spatial distribution is highly heterogeneous.

\bibliographystyle{IEEEtran}
\bibliography{IEEEabrv,reference}

\end{document}